# Biomechanics of Collective Cell Migration in Cancer Progression– Experimental and Computational Methods

Catalina-Paula Spatarelu+, Hao Zhang+, Dung Trung Nguyen+, Xinyue Han, Ruchuan Liu, Qiaohang Guo, Jacob Notbohm, Jing Fan, Liyu Liu*, Zi Chen*

+ These authors contributed equally

*Emails: zi.chen@dartmouth.edu; phyliurc@cqu.edu.cn.

Mailing Address: Thayer School of Engineering, 14 Engineering Drive, Hanover, NH, 03755, USA

**Keywords:** cell mechanics, cancer biophysics, unjamming, collective cell dynamics, cancer invasion, epithelial-mesenchymal transition

**Abstract**

Cell migration is essential for regulating many biological processes in physiological or pathological conditions, including embryonic development and cancer invasion. In vitro and in silico studies suggest that collective cell migration is associated with some biomechanical particularities, such as restructuring of extracellular matrix (ECM), stress and force distribution profiles, and reorganization of cytoskeleton. Therefore, the phenomenon could be understood by an in-depth study of cells' behavior determinants, including but not limited to mechanical cues from the environment and from fellow "travelers".

This review article aims to cover the recent development of experimental and computational methods for studying the biomechanics of collective cell migration during cancer progression and invasion. We also summarized the tested hypotheses regarding the mechanism underlying collective cell migration enabled by these methods. Together, the paper enables a broad overview on the methods and tools currently available to unravel the biophysical mechanisms pertinent to cell collective migration, as well as providing perspectives on future development towards eventually deciphering the key mechanisms behind the most lethal feature of cancer.

## I. Introduction

While single cell migration has been well studied, the mechanism of collective migration still remains elusive. Cell collectives are not simply a collection of single cells, and as such, collective migration involves much more mechanically and dynamically complex intra and inter activities[1,2]. Collective cell migration involves a group of cells travelling together under certain patterns with a degree of coordination among the movement of each individual[3]. Despite the involvement of such phenomenon in several physiological processes, the underlying biophysical mechanisms are far from being thoroughly understood.

During embryogenesis, certain phenomena such as neural tube closing, ventral closure or eyelid closure[4–6] rely on neighboring epithelial cells collectively migrating from stem cell niches and proliferative zones to the target functioning area[7], and consequently forming a continuous monolayer[8]. Different levels of development of embryos show distinct patterns of collective migration. For example, early embryos employ sheet migration where cells remain tight and close while moving forward together[7,9]. Tracheal system shows branching morphogenesis, as it generates a limited number of tip cells and forms elaborate cellular structures[3,7]. Border cells in *Drosophila* ovary migrate as a free group consisting of leaders (cells receiving constitutive epidermal growth factor (EGF) receptor signaling) and followers[3,7,10]. The lateral line in zebrafish has a slug-like migration[7,11]. Mechanical forces in all these processes depend not only on actomyosin contractions which form a purse-string structure[5,8] but also on cell crawling driven by lamellipodia and filopodia extrusions[12,13].

Collective migration is involved in different stages of wound healing as well. During the generation of epithelial monolayers, leader cells are believed to drag follower cells to migrate, depending on the cell-substrate traction forces localized to the leading edge[14,15]. However, evidence, supporting that not leader cells but rather an integrative process of transforming local forces into a general tension in the monolayer by each cell, was presented as well[16,17]. In re-epithelialization of epidermal cells, keratinocytes migrate collectively across the provisional wound bed which is filled with ECM. This sheet migration results in epidermal wound closure[15,18,19]. In the regeneration of endothelial cells, collective strands of endothelial cells migrate as vascular sprouting with a leading sprout, and penetrate the provisional wound matrix to regenerate vessels[3,7,20]. Researchers have been working on modeling the collective migration during wound healing using continuum model and free boundary system[21]. However, the specific mechanism(s) and the determinants of migration and the migration mode still remain unclear.

Cancer invasion, mostly for epithelial-like tumor cells[7], displays many hallmarks of collective cell migration. Firstly, cell-cell junctions, suggesting cell-cell couplings, are observed among cancer cells[3,22]. Secondly, cell-cell adhesion molecules such as E-cadherin and homophilic cell-cell adhesion receptors are expressed within invasive zones[3,23]. Thirdly, tumor cells display structural ECM remodeling on early movement, suggesting their capability of morphogenesis[3]. Lastly, cancer cells migrate from invasive zones to the surrounding matrix as chains or sprouts[7,24]. Cancers that display collective invasion include partially de-differentiated forms of rhabdomyosarcoma, oral squamous cell carcinoma, colorectal carcinoma, melanoma and breast cancer[19,23–25]. While the molecular mechanism of cancer collective invasion is not completely understood, several interesting findings have gradually been made. Understanding the biomechanical determinants of cancer cells before or during collective migration can help identify medical targets. Investigating diseases from a biomechanical point of view has been an important tool for elucidating pathophysiology and pathogenesis of a variety of ailments[26], and has helped in developing diagnosis and therapeutic strategies[26,27]. These have allowed for the recognition of several differences between cancer cells and corresponding normal cells, from a biomechanical perspective. Some notable distinct biomechanical characteristics of tumor cells, at a single-cell level, include their higher stretchability[26], greater softness[28] and loss of stiffness sensing abilities in the case of some malignant cells[28], that might have implications on their collective behavior as well.

In epithelial cancers, cells migrate collectively from the main mass to surrounding tissues[29]. In metastasis of other cancer types, besides those of epithelial to mesenchymal transition (EMT) and migratory phenotypes[30], cells also exhibit collective migration to spread to other tissues[29,31]. As a result, controlling collective migration might be an effective strategy of dealing with cancer spread.

Different from the review by Hakim and Silberzan[32], this review paper focuses on both the experimental and computational methods employed to study collective cell migration, highlighting some of the key discoveries that add to our understanding of the phenomenon. A brief overview of the mechanistic basis is given in section II, followed by the review of most commonly used experimental and computational methods, as well as the identification of the unresolved questions regarding collective cell migration during cancer progression and invasion. A more detailed review on the cell-matrix related mechanobiological changes during cell migration can be found elsewhere[33].

## II. Mechanisms of Collective Cell Migration

### Cues for Collective Cell Migration

Before going into the mechanism of collective cell migration, it is important to identify the cues that trigger the process. Cues for solitary cell migration are well studied, as it has been shown that cells can move along a gradient of chemoattractant (chemotaxis), adhesive substrate (haptoaxis) or rigidity (duroaxis), or move in response to electric fields or topography[34]. However, more complex molecular pathways are often involved in converting external guiding-factors into migration events for cell collectives. For example, the recent work by Barriga et al. has shown that a change in substrate stiffness can trigger collective cell migration[35], and in this pathway, TWIST1 is the key mechano-mediator--upon sensing the high matrix stiffness, TWIST1 is released from its binding partner and translocated to the nucleus to induce constitutive TWIST1 nuclear translocation and EMT[36]. While extrinsic factors such as chemoattractants, electric fields, or topography can single-handedly dominate and guide the migration of single cells, collective cell migration requires both the integration of these regulators among the collective and a cohesive response[34,37,38]. Although many methods to study the chemical and electrical cues for migration have been developed, studying mechanical cues other than stiffness remain challenging due to a lack of techniques to effectively identify and quantify the mechanical signals. In addition to the external chemical, intrinsic factors among the cell collective also play a role by potentially affecting migration direction, velocity, and capability. Within the migrating cell collective, integrated intrinsic cell-cell interactions like mechanocoupling and polarization guide the overall movement, but cells can also sense and respond to extrinsic cues on the individual-level[34,35]. Traction force, which directs cell migration, is regulated by cell shapes and substrate stiffness[39,40]. Differences in expression profiles between cells in the same migrating collective have been identified, and consequently support the hypothesis of leader-follower cell dynamics[41].

**Leader-Followers Cell Dynamics**

"Swarm-like" behavior of cancer cells proposed by Deisboeck et al. presents the paradigm of tumors as units with multiple individuals (cells) among which we find capabilities of leadership, conflict and cooperation[42]. While in the case of single-cell migration one could discuss the polarization of a cell to yield a leading edge and a trailing one, in the case of collective cell migration, the occurrence of leader cells and follower cells has been observed to coordinate a collectively polarized group of cells[43]. The occurrence of the polarization is assumed to be caused by the change of cadherin-mediated contacts due to pulling forces within the monolayer[1]. This pressure is hypothesized to encourage actin polymerization at the individual cell front, previously described to enable a push forward of the leading edge of leader cells [44]. In addition, it has been suggested that the collective migration of metastatic cells act in a cooperative manner by exchanging the leadership of the front cells[45]. However, the collective movement is sustained not only by the existence of leader cells and follower ones, but also by the interactions through the collective, that maintain mechanocoupling and guidance. Cell-cell junctions contribute to supracellular adhesion and mechanocoupling sensing and interpretation of external cues, such as soluble factors or ECM topology[43]. Their connections are realized through cadherin adhesion proteins, desmosomal proteins, tight junctions, gap junctions and interactions between immunoglobulin family proteins. Cadherins exert signals along cell-cell junctions, while E-cadherin and N-cadherin levels are often used as markers for cell motility[29,46]. Mechanosensitive molecules, such as vinculin and filamin, play an important role by changing their conformations in response to forces transmitted at cell-cell junctions triggering signaling events[3,47]. Inhibition of the integrin-vinculin-talin complex disturbs the mechanosensing, and consequently aborts the collective cell migration[35]. Moreover, actomyosin was shown to accumulate at cell-cell adhesion locations inducing an inner polarization of leader cells, which increases collective migration by creating a permanently tensioned monolayer[48]. Because of the difference in traction forces experienced by leader and followers, their interaction has been deemed as resembling a "tug war"[49], in which the stronger leader cells prevail; however, success is credited partially to the existence of actin cables along the leading edge that help maintain the integrity of cell-cell adhesions [29], or to the alignment of forces with the velocity of the collective[50].

Figure 1 provides a schematic representation of leader-follower cell dynamics during collective migration, together with key factors involved in the migration mechanism. The selection of the leader cells in cell collectives is done by a combination of synergistic mechanisms that aim to maintain a polarized state of the migrating layer consisting of a leading edge and a trailing one, while ensuring the integrity of cell-cell contacts and the stability of the tissue geometry[29]. Leader cells have been observed to manifest different phenotypes than follower cells, being distinguished by prominent pseudopodia aimed in the direction of migration, as portrayed in the schematic representation in Figure 1.

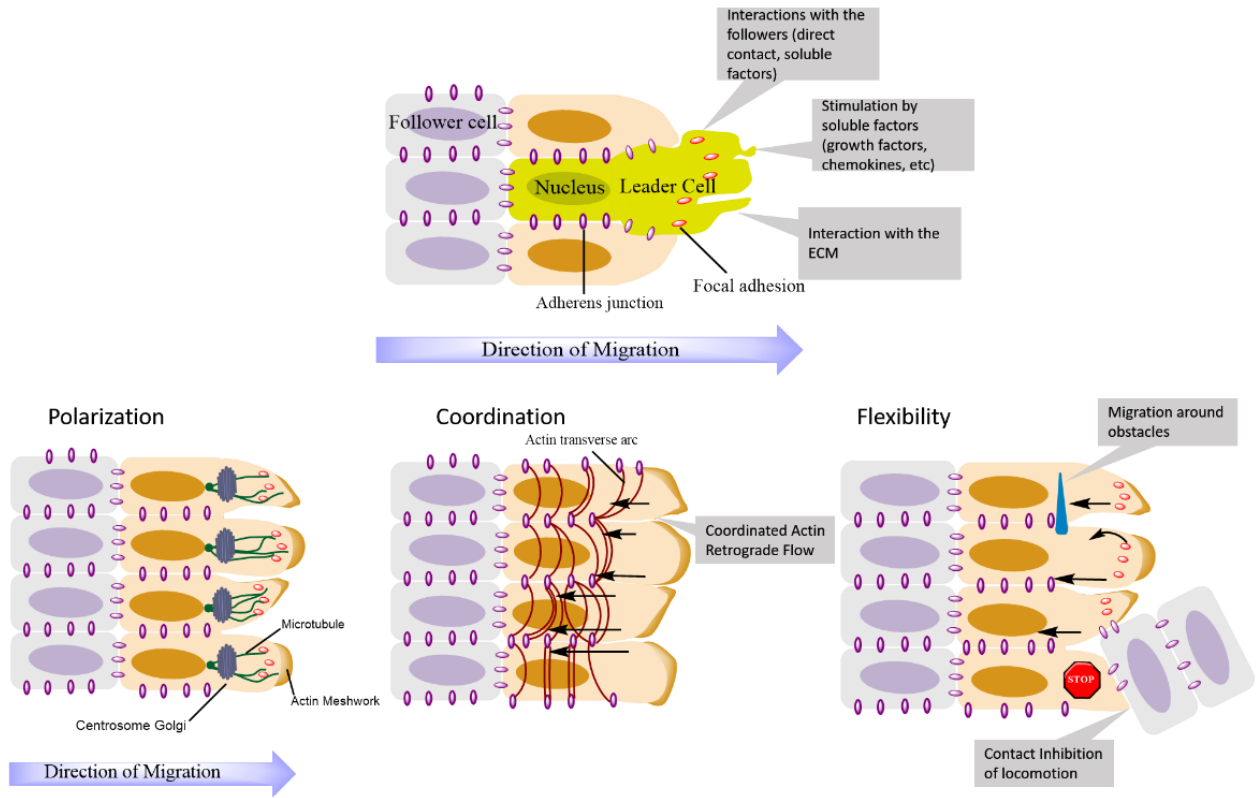

***Figure 1*** *Top: Representation of the emergence of leader and follower cells within a cell cluster; Bottom: 3 key elements of collective cell migration represented with their main components: a. polarization of cells that leads to the orientation of the cytoskeleton; b. coordination of actin dynamics between neighboring cells through actin cables; c. flexibility or dynamic rearrangement of cell-cell contacts due to the retrograde flow of adherens junctions. Redrawn with permission from [29]. Copyright 2016, Springer Nature.*

The terminology of "leaders" has been, however, questioned as being deceiving in describing cells that possess the guiding or steering capabilities within a group, as their positioning does not need to be at the front of the collective[51]. Furthermore, the existence of a stress coordination among cells[17] and the occurrence of plithotaxis among cell collectives, based on mechanical cell-cell contact[52], emphasize that the leader-follower dynamics might be just a subset of a bigger picture to describe the modes of collective migration. The following sections will explore this idea, describing the shift from the biochemical-based view of collective migration at a cellular level to a more integrative one, which involves factors concerning the mechanical interactions within the collective as a whole.

**Glass-like Behavior of Cells**

Introduced in 2001 by Fabry et al.[53], the analogy of cell behavior with that of soft glassy materials (SGM) is a concept that has attracted considerable interest from biomechanics researchers. SGMs include diverse substances such as foams, colloids, emulsions or slurries, and are characterized by their numerous soft elements, aggregated with one another by means of weak interactions and existing away from thermodynamic equilibrium[54]. Fabry et al.'s study looked at the dynamics of the cytoskeleton and proposed that the cytoskeleton might undergo aging, rejuvenation and remodeling events in a similar way to soft glassy materials. This led to the proposition that cells be included in the list of SGMs. Angelini et al. expanded the analogy to compare the fragility analysis of cell layers during collective migration and atomic and molecular glasses[55]. Moreover, a study on individual cells, showed that compression results in a stiffening of the cytoskeleton to resist it and a decrease in the relaxation time[56]. This response of the intracellular space is reminiscent of a repulsive colloidal suspension approaching a glass transition. The projection of this glass-like behavior in cell collectives as well as within the cell is supported by evidence from studies that described structural rearrangements[57,58], kinetic phase transitions shown to occur in cell monolayers[59], and glass-like characteristics during wound closure[60].

In the case of inert soft condensed matter, the increase in the size of cell clusters causes a gradual decrease in velocity, which leads to a kinetic arrest phenomena[54]. However, the internal structure remains disordered, and the jamming transition requires fine tuning of temperature, density, and shear stress[61]. Similarly, in a cell collective, with increasing cell density by proliferation, the velocity fields observed become more heterogeneous, reminiscent of a liquid-like to solid-like transition[62].

**Jamming/Unjamming Transitions**

Further expanding on the existence of solid-like (caged or "jammed") first proposed by Lenormand et al. in 2007[63] and liquid-like behaviors of cell collectives ("unjammed"), the existence of a jamming/unjamming transition phenomenon was proposed for epithelial cell collectives[59,64,65]. A hypothetical diagram of the transition is shown in figure 2. In the jammed state, cells are confined in such packing geometry that the thermal fluctuations are insufficient to drive local structural rearrangement[66]. The jamming phenomenon is similar to the one present in inert matter[67,68], with the particularity that cell collectives contain active particles with a motility of their own and that jamming transitions are encountered at packing densities of 1, such as in confluent monolayers[65].

Several parameters are considered to have a role in the unjamming/jamming transition, such as cell density, cell-cell adhesion, cell motility, substrate mechanical properties such as stiffness, cell stresses and their distributions[69]. However, the main determinants of a potential jamming/unjamming phase diagram such as the one in Figure 2, are still being debated, potentially due to the intricacy of the relations between multiple factors. Some authors argue that geometry and the depart from an equilibrium geometry are drivers of unjamming in epithelial cells[70], while others view the increase in mechanical coupling as a prerequisite[71]. The density of the cells within a monolayer has been considered as a factor[55], but other works seem to debunk that idea[72,73]. While these viewpoints might seem contradictory, the reader is advised to regard the main results presented within this review as complementary, alas potentially incomplete. In a similar fashion, the phase diagram represented here, or in other papers regarding the jamming/unjamming phenomenon in cells, is likely only a partial representation of the reality.

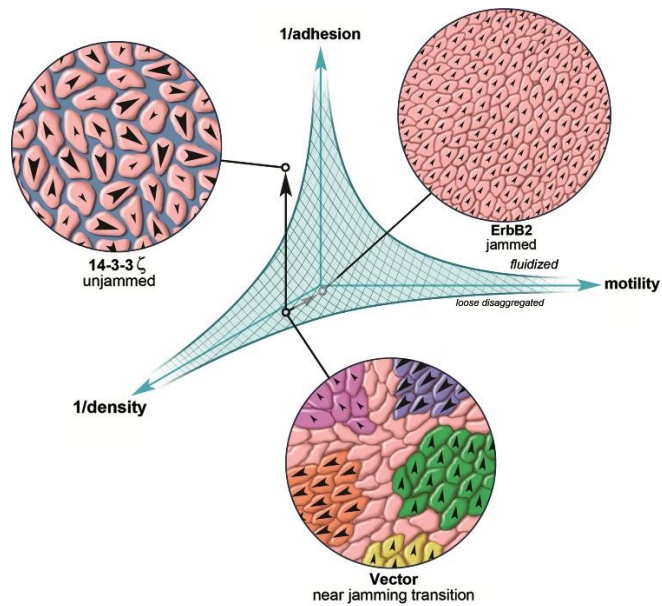

**Figure 2**. *Hypothetical phase diagram of the jamming/unjamming transition. The shaded area represents the jammed state, which is influenced by the density of cells, the motility of cells in the monolayer and the cell-cell adhesion strength. Reproduced with permission from[69]. Copyright 2013 International Society of Differentiation. Published by Elsevier Inc.*

Cellular jamming and unjamming have been observed in experiments by Park et al. with primary epithelial cells from asthmatic and non-asthmatic donors. They showed that after undergoing an unjamming transition caused by external mechanical compression, non-asthmatic tissue would transition back to a jammed state in around 6 days, while their asthmatic counterparts exhibited a delay in reaching this state, taking up to 14 days to become jammed[74]. In this context, the unjamming transition in cancerous tissue and other pathologies might be seen as an aberrant behavior of epithelial cells that disturbs the idle, jammed state of these cells[75]. Compression-induced unjamming has been observed in human bronchial epithelial cell layer, and the process is associated with certain changes in cell shape including apical flattening, constriction of apical diameter, cell elongation, and more shape variability. However, whether apical actomycin constriction leads to unjamming remains unclear[66]. Unlike inert matter whose boundary conditions do not influence the internal parameters of the system, a wound that changes boundary conditions can slowly lead to decreased density and consequently unjamming[76]. The jamming/unjamming transition is also affected by the cell type: while proliferation of epithelial cells leads to migration arrest, mesenchymal cells that proliferate slower and migrate faster can impede cell jamming[77].The underlying mechanism of the transition is proposed to be based on mechanical energy barriers to cell rearrangements, contributed by cell-cell junctions, contractile energies and adhesion energies. As the energy barriers go up, such as due to smaller adhesive with respect to contractile contributions, the systems jams. The opposite happens when the adhesive energy overwhelms the contractile one, with cells becoming fluidized and unjammed[78].

The response of mesenchymal cells under ECM confinement provides evidence for cell jamming being an essential component for the emergence of collective cell migration modes[79]. Thus, mesenchymal cells were shown to switch from a single-cell migration mode to a collective one when the microenvironment confinement was large enough to cause cells to jam and reach a large density[79]. While collective cell migration is a property of normal epithelial and endothelial collectives[7], observing the phenomenon in mesenchymal cells might have implications on the approach in dealing with cancer metastasis. Given the increased division rates in cancerous tissue that might lead to fluidization, tumors could be inherently prone to unjamming[65]. This, together with the new knowledge that EMT is unessential for the development of secondary tumors[80,81], supports the jamming/unjamming transition as a good contender for a triggering mechanism. However, as stated before, current progress in the field does not allow for a definite set of parameters to be accepted as key players in the involved phenomena. Moreover, observations made for a certain cell type, or tissue type, might not necessarily hold true regarding another.

### III. Experimental methods

An attempt to fully describe the biomechanical mechanisms underlying collective cell migration or cancer invasion will unquestionably involve experiments observing these phenomena in a controlled manner. Based mostly on cell culture methods, experimental tools and techniques that were and continue to be used in the study of collective cell migration could be regarded as 2D techniques or 3D techniques, with subdivisions for both large categories. This section reviews some of the most commonly used methods that have added to our knowledge of biomechanics of collective cell migration in general and in the case of cancer in particular.

**2D Cell Culture Methods**

*Migration assays*

The most commonly used and known migration assay is the Boyden chamber, consisting of a transwell that allows the cells and a chemoattractant agent to be separated by a porous membrane through which cells can migrate[82]. Although the setup is fairly easy and accessible, and this method has been used to study the effect of several biochemical cues on invasive and migratory behavior of cancer cells [83–85] , the method lacks control over geometrical factors and does not offer the possibility of visually inspecting the migration process, thus limiting its applicability. Therefore, the method might be better suited for determining single cell invasion or migration characteristics rather than collective behavior.

To address some of these shortcomings, cell exclusion assays were proposed as an alternative. An example focusing on collective cell behavior is the setup used by Nyegaard et al.[86] that has been shown to give reproducible results, with high versatility towards ECM types and compatible cells. The principle behind it is to place a barrier before seeding cells and by removing it to allow the cells to migrate into the newly formed void. A drawback of this method was its inability to exclude the effect of proliferation, but by using the lineage-tracing vital stain, one can distinguish migration from proliferation[87]. However, in the context of collective cell migration, this technique can be improved using methods of lithography, to obtain highly regular substrates and representative micro-topographies[88–90].

Microstencils obtained by soft lithography allow the confluent cell monolayers to migrate. For instance, Poujade et al. used this technique to validate the triggering of collective motility by the availability of newer space to migrate, rather than by an injury of a previously confluent monolayer[23]. Petitjean et al. used the same method to study the collective migration of epithelial sheets, particularly identifying the different phenotype of emerging leader cells that were organizing as "finger" structures through particle image velocimetry and particle tracking[22]. This study built up to the conclusions by Poujade et al. that employed similar techniques to validate the triggering of collective motility by the availability of newer space to migrate, rather than by an injury of a previously confluent monolayer[23].

Notbohm et al. used micropatterned masks to investigate the oscillatory behavior of epithelial confined monolayers, and a mechanochemical feedback mechanism was proposed as an explanation for the periodic fluidization and stiffening of cell monolayer observed[50]. In the context of both single and collective cell migration behaviors, topography and anisotropy were shown to influence migration patterns and their orientations, as shown by studies employing micropatterned topographies[93].

*Wound-healing Assays*

Unlike migration assays, wound-healing assays usually imply a form of cell injury that results in a cell depleted area[94]. In a 2D assay, this is usually obtained by letting cells reach confluence and injure the monolayer through various ways: scratching, stamping, thermal or electric wounding or laser ablation.

Due to the ease to set up, wound healing assays are useful in the study of collective cell migration in conjunction with various microscopy techniques[95]. Interestingly, studies focusing on wound-closing have shown that the process is independent of proliferation but is correlated with cell area and persistence[96], which suggest that the cell density is not a determinant in initiating collective cell migration in this context[97]. Biochemical pathways of collective migration have been investigated, with Nobes et al. looking at individual contributions of Rho, Rac, and Cdc42 to cell movement in the context of wound healing. Their work revealed a mechanism based on the collaboration of small GTPases Rho, Rac, Cdc42 and Ras, where the first is required to maintain cell adhesion, the second is essential for the protrusion of lamellipodia, while Cdc42 regulates cell polarity and Ras is involved in controlling focal adhesion and stress fiber turnover[98]. Additionally, hypoxia and cadherin-22 have been shown to co-localize in human glioblastoma and promote collective cell migration by means of cell-cell adhesions[46].

As collective migration heavily relies on cooperation of numerous individuals, the role of cell-cell interactions and their dynamics during migration is a key element to inform a better understanding of the mechanisms involved. Tamada et al. used a circular wound model obtained by laser ablation to study the cell-cell adhesions rearrangement during collective migration, as well as to reveal two mechanisms of closing the wound: the actomyosin assembly and contraction that acts on apical edges, together with the lamellipodial protrusions into the cell depleted area[99]. De Pascalis et al. recently studied glial interfilaments and their influence on astrocytes' collective migration speed, direction and persistence. Their study proved that the interfilament network controlled the traction forces distribution in the migrating sheet, as well as holding an important role in the organization of actin cytoskeleton and focal adhesions to maintain the structural and functional integrity of the collective[100].

**3D Cell Culture Methods**

*Spheroids*

As by far the most widespread method for studying collective cell migration in three-dimensional culture, the cancer spheroid model can incorporate factors that were not accounted for in the 2D models. Spheroids are constituted by cells, either primary or from cell lines, aggregated in a 3D scaffold, usually from ECM proteins (collagen, fibrin, Matrigel, etc.)[101]. The main categories of these cancer spheroids are multicellular tumor spheroids, tumorspheres, tissue-derived tumorspheres and organotypic multicellular spheres[101]. Their compositions and applications usually differ: multicellular tumor spheroids are obtained from cell suspensions and used primarily to assess cell migration through ECM; tumorspheres are usually derived from stem cancer cells and employed for studying cancer stemness; tissue-derived tumor spheres and organotypic spheres are derived from primary cancer tissue and are often used to screen drugs or characterize a patient's tumor[102]. Several key factors were studied in the context of cell migration in tumor spheroids, such as matrix density and stiffness[103], ECM reorganization by means of proteases such as MMP, cell-cell adhesion and their dynamics, microenvironment stresses[104]. For example, Labernadie et al. examined the heterophilic E-cadherin/N-cadherin junction in the tumor spheroids made up of cancer cells and cancer-associated fibroblasts to show that these heterotypic junctions serve as force transmitters and mechanotransduction triggers that lead to collective cell migration and invasion[105]. Malignant breast cancer cells MDA MB 231 were cocultured in spheroids with MCF-10A epithelial cells by Carey et al. to study the role of leader cells observed in collective cell migration patterns[106]. Their findings suggest that malignant cells can "recruit" follower cells from epithelial, non-malignant cells within the tissue, by cell-contractility and proteolysis-dependent matrix remodeling. Also notable are the results of Haeger et al. that studied the response of MV3 melanoma and HT1080 fibrosarcoma cells cultured in spheroids of varying ECM densities. Their findings indicate a plastic response of the cells, that involved the shift from single-cell to collective invasive strategies given various degrees of confinement[79].

*Wound-healing Assays*

Although less used than the tumor spheroid model, wound-healing assays have been developed for studying collective cell behavior in 3D in vitro. The constructs used in 3D wound healing assays usually consist of layered structures, combining ECM components such as collagen hydrogels and cells[94]. Topman et al. used such an assay with a transparent hyaluronic-acid based hydrogel to calculate the migration rate and directionality of DAPI-stained cells in vitro[107]. As studies have shown that the behavior of cells differs from 2D to 3D environments[108,109], the analysis of collective cell migration in 3D environments and the processing of such data is of uttermost importance in revealing the mechanobiological mechanisms involved. Particularly, a study by Hakkinen et al. looked at the influence of 2D vs 3D tissue cultures on morphology, adhesion, single-cell migration and cytoskeleton for human fibroblasts grown in 4 different ECMs[108]. Their work revealed the differences in characteristics such as cell morphology, migration direction and rate, adhesion to ECM and actin stresses for both 2D vs 3D environments, as well as between the different ECMs used. These findings strengthen the need for faithful reproduction of tumor realities within the models used for in vitro testing, and justify the use of 3D models, such as the ones mentioned above (organoids, 3D wound healing assays) or microfluidics-enabled ones.

**Microfluidics-based Methods**

Advances in microfabrication methods made possible the existence of microfluidic devices that mimic the tumor microenvironment and facilitated a better understanding of the micromechanical cues that regulate collective cell behavior[110]. Microfluidics-based methods have the advantage of single-cell resolution and tailoring ability, allowing for the study of the complex biomechanical interactions that take place during cell migration, from single-cell migration to collective behavior modes[111]. Apart from biochemical cues, cell migration behaviors are informed by the surrounding tissue and ECM properties the reproduction of which are somewhat restricted in the macroscopic cell culture methods[110,111] previously discussed. Some of the common soft-lithography and microfluidics-based methods are shown in Figures 4 and 5.

Studies that proved the cytoskeleton organization[112], focal adhesion assembly[113] or cell alignment[114] change as a mechanotransduction response to microtopography, stand to show how essential it is to include these type of factors into the study of cell migration. This knowledge might help inform better in vitro prototypes that elucidate the role of mechanical cues in migratory mode. Wong et al. studied the behavior of cells at the tumor invasion front by employing micropillars to disrupt the cell-cell contacts periodically to enhance individual scattering of cells in a confluent layer[115]. Liu et al. developed a microfabricated chip with micropillar arrays as a 3D landscape to investigate the invasiveness of metastatic cells quantitatively[116].

Recently, Cui et al. developed a microfluidic device with multiple cell collection microchambers to study the morphology, cytoskeletal structure, and migration of transepithelial-migrated cells with high specificity, so a fully covered epithelial layer is no longer required for migration[117].

Trending towards mimicking tumor microenvironments more closely, cancer models have been developed by means of microfluidic devices. These methods are being employed as a versatile tool to study the influence of several factors on the migration behavior of cells[111,118]. The roles of biochemical cues such as AURKA kinase[119], chemoattractants[120], or even mechanical characteristics of the environment such as confinement[121] and topology[122] have been investigated using such methods. Furthermore, high-throughput microfluidic chip technique has also been utilized to explore tumor-stroma interactions in a 3D microenvironment[123]. An important aspect of the tumor microenvironment that might be challenging to capture using the methods described in previous sections is its heterogeneous nature. As previously pointed out in this review, the mechanical environment is a significant determinant of cell migration and any heterogeneity that appears might be responsible for behaviors observed in in vivo cancer progression[124,125]. To that end, an interesting approach by Alobaidi et al., consists in a Diskoid in Geometrically Patterned ECM or "DIGME", a mechanical-based method that allows for the control of the shape, microstructure and heterogeneity of tumor organoids[126].

At the later grade of carcinoma tumor, collagen fibers aligned with tumor cells have been observed. In order to investigate the impact of collagen fibers on the tumor cell invasion, Han et al. designed a sandwiched ECM in a microfluidic chip to create collagen fibers perpendicular to the interface of collagen gel and Matrigel (Figure 3a and 3b). In such a heterogeneous system, the oriented collagen fibers facilitate the invasion of breast cancer cells (MDA-MB-231) into the 100% Matrigel layer as fast as in 96 hours (Figure 3c), while in a homogeneous collagen gel the cells were unable to invade the Matrigel. In addition, in 144 hours, the front cells successfully break through the entire Matrigel layer, indicating the influence of oriented collagen fibers goes much further than their penetration lengths. The results demonstrate the essential role of collagen fiber orientations in guiding metastatic cell intravasation and assisting the breakage of basement membrane before entering the circulation systems[127].

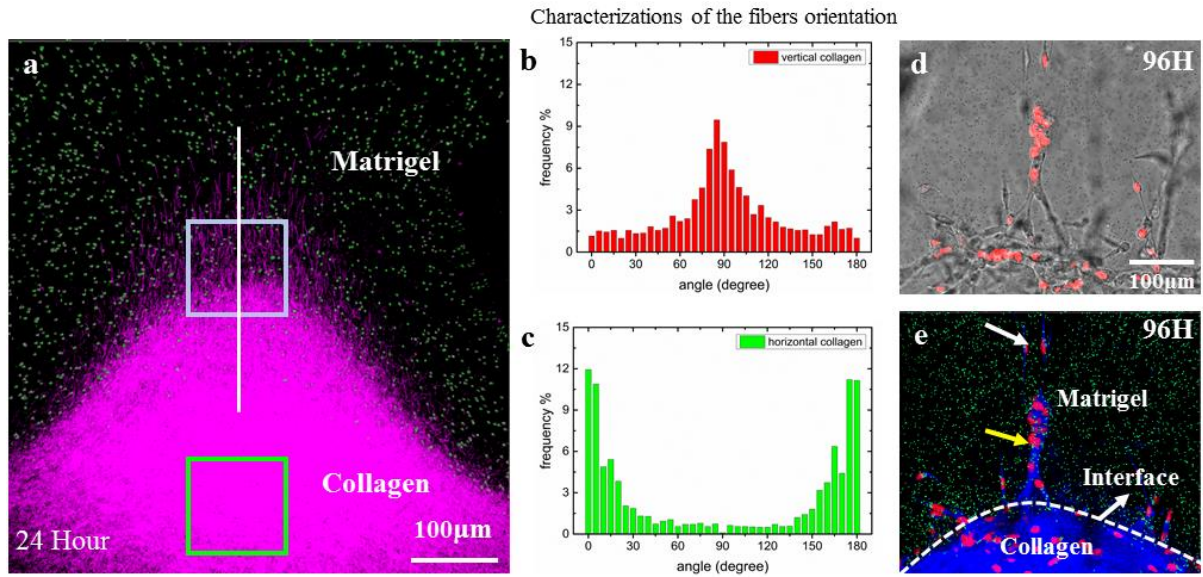

**Figure 3.** Oriented collagen fibers direct the metastatic cell invasion into Matrigel. a. The 3D confocal image reconstruction via image stacking (top view) shows the Matrigel/collagen composite ECM and their interface in three dimensions. b. The collagen fibers near the interface region possess vertical orientations (red box). c. The collagen fibers in the centered region possess horizontal orientations (green box). d. The bright-field images showing snapshots of invading cells at 96 hours. e. The corresponding fluorescent images combined with reflective mode of (d), which show the Matrigel region with green beads embedded, the collagen region (blue), and the nuclei of invading cells (red). It isclear that, at the 96th hour, guided by oriented collagen fibers, the cells aggregated and strongly invaded into the rigid Matrigel region in single-stream forms. The field of view is the same for (d) and (e). Reproduced with permission from[127].Copyright 2016, Proceedings of National Academy of Sciences.

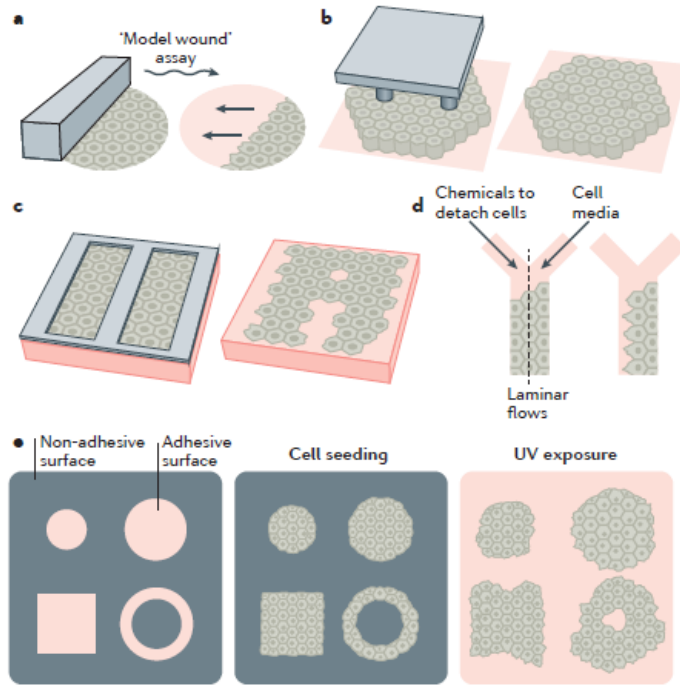

**Figure 4.** Example of microfluidics-based methods to study cell migration. a) Model-Wound when a microfabricated barrier or mark physically confines cells and is removed to reveal space for monolayers to migrate onto; b) Micropillars; c) Microstencils; d) Microfluidic device for chemically excluding cells; e) UV-cleavable mask that allows for the confinement of cells and their subsequent release by irradiation. Reproduced with permission from [128]; Copyright 2017 Springer Nature.

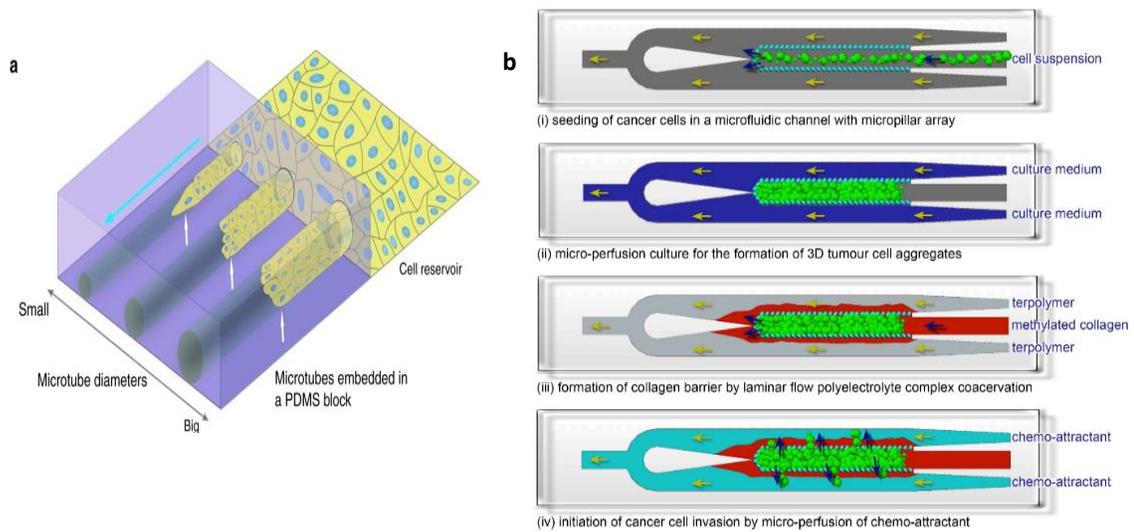

**Figure 5.** a. Schematic of MDCK cells feeding a PDMS block with varying sized microtubes; Reprinted from[121]; b. Schematic of a cancer cell migration assay with a microfluidic device and the steps of studying migration in a 3D environment (**i**) Seeding of cancer cells (**ii**) Micro-perfusion culture to allow 3D tumor aggregates to form (**iii**) Formation of a collagen barrier around the cellular aggregate by coacervation of positively charged collagen and negatively charged synthetic terpolymer (**iv**) Migration of cancer cell by micro-perfusion of chemoattractant; Reproduced with permission from [120]. Copyright 2018 Bioengineering.

**Measuring Velocities during Collective Cell Migration**

Even in the leader-follower type of organizations for collective cell migration, the front of the migration is not the only motion in monolayers. The bulk of the cell monolayer often experiences swirling that causes 'chaotic' motion within the confluent layer[74]. The advent of techniques such as particle image velocimetry allows for the characterization of these collective movements in terms of directionality and velocity[92]. Heat maps of the velocity fields and displacements in confluent layers, such as the ones shown in Figure 6, were used to study the influence of cell density that is believed to lead to a decrease in cellular rearrangements yielding a jammed state of the epithelia, as well as the effects of cell-cell adhesion or cortical tension variants.

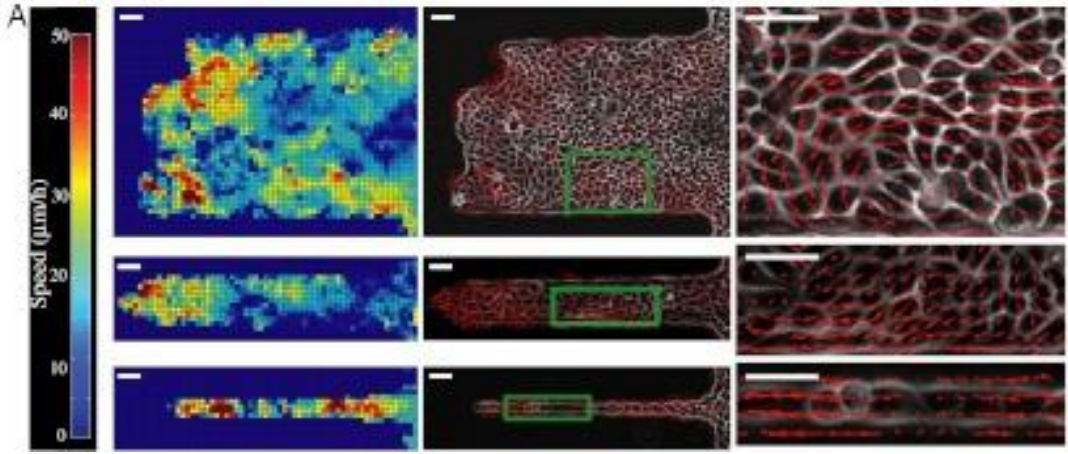

**Figure 6**. PIV analysis of epithelial sheets migrating on fibronectin strips of varying widths showing the spatial distribution of velocity fields as a heat map (right); A change in the collective behavior is recorded for the larger channels that move by swirling motions compared to the contraction-relaxation based mode experienced through the narrow channels. Reproduced with permission from [129]. Copyright 2012 Proceedings of National Academy of Sciences.

**Measuring Forces during Collective Cell Migration**

Another fundamental component towards understanding the biomechanics of collective cell migration is the characterization of forces during the collective behavior. Multiple techniques are employed to infer stresses and forces for both in vitro experiments, as well as in vivo, looking at scales between molecular and tissue level[78]. One commonly used method consists of mechanical traction force microscopy (TFM), which uses the deformation of soft substrates, embedded with fluorescent beads and knowledge of its mechanical properties to infer the traction applied by the cells to the substrate[130]. Dembo and Wang first fully developed the method to manifest and quantify the tractions using TFM, and used the method to demonstrate that the ability lamellipodium to generate intense traction stress in 3T3 fibroblast[131]. Butler et al. soon improved the method so tractions can be more easily solved from displacements[132], and Trepat et al. applied this method to collective cell migration[16].

A similar principle is based seeding cells on a substrate of soft micropillars and inferring the tractions by the pillar deformations. For example, Du Roure et al. used this setup to demonstrate that the highest average traction of a migrating epithelial monolayer is localized at the edge of the layer and oriented towards the bulk of the cell layer[133]. Figure 7 shows the results of Reffay et al., who used this method to make similar observations in finger-like formations during the collective migration of MDCK cells.

Stresses between cells can be computed from the tractions and the principle of equilibrium[17,134,135]. These methods give the tractions and stresses that together bring about collective migration. Details on the relationship between force and motion are still being explored. In their work using computational modelling, Yang et al., explore the possibility of integrating TFM measurements and mechanical inference to account for the non-static character of cells[136]. Another recent finding is that stress anisotropy produces alignment between local maximal stress within a cell monolayer and the local migration velocity of the cells[17], phenomenon recognized as "phlithotaxis". Other works have shown a quantitative relation between shear stress and cell-alignment[137] and the increase in strain rates as a precursor of coordinated motion[138]. Integrating these effects, Zaritsky et al. proposed that the mechanism of locally induced coordinated movement relies on the leader cells inducing normal strain on rear neighboring cells, and shear strain on adjacent cells[139].

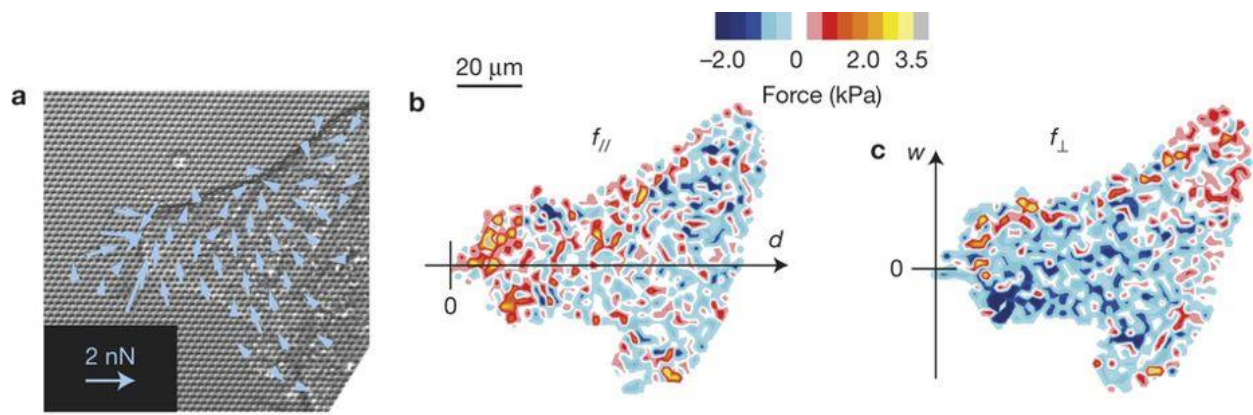

**Figure 7**. Traction force maps obtained in a finger-like formation of MDCK cells migrating on soft micropillars: (a) direction of main forces (b) intensity of the longitudinal direction (c) intensity on the transverse direction. Reproduced with permission from [140]. Copyright 2014 Springer Nature.

Forces can also be assessed by using FRET as a tension sensor by incorporating the FRET probes within proteins from the cytoskeleton, cell-cell junctions or cell-ECM adhesion proteins. For instance, Sarangi et al. combined micropillar substrates with a FRET molecular tension sensor that allowed them to map both the traction forces of each focal adhesion cluster and the molecular tensions experienced within the vinculin molecules in those clusters, showing a strong spatiotemporal correlation between the two[141].

Although 2D and 3D assays are helpful in investigating effects of certain factors on collective migration, or the biomechanics of cell collectives during the phenomena, the in vivo reality might be different, much as 2D behaviors differ from 3D ones for various types of cells being studied. In this context, it is natural to envision methods to look at the phenomena in vivo. With the ease of fluorescence labeling, either genetically (through FRET sensors or simply fluorescent proteins to allow for detection) or by live-staining methods, and microscopy techniques improving, the hurdle of in vivo experiments strongly lays on the side of data processing rather than data acquisition. Cliffe et. al developed a processing toolbox for investigating collective migration in drosophila[142], while other studies in vivo have focused on zebrafish[143] or Xenopus laevis[144] embryos. However, establishing FRET probes for in vivo models remains challenging, as Eder et al. discuss in their paper regarding FRET E-cadherin tension sensors in a *Drosophila melanogaster* model[145]. Moreover, studying collective migration in vivo in the case of cancer brings another set of challenges, as the number of cells is increased dramatically and imaging procedures are limited by time and field of view[146]. Therefore, aspects such as the significance of stress-guided migration patterns or a precise mapping of the forces involved in vivo remain open questions for the moment. An approach to mitigate this type of shortcoming is to produce computational models, versatile and readily available to simulate in vivo behaviors, the highlights of which are described in the following section.

*FRET biosensors*

FRET (Foster Resonance Energy Transfer) biosensors are based on the interactions of the electromagnetical fields of two fluorescent molecules with overlapping excitation and emission spectra, one of which is deemed the donor and the other the acceptor[147]. The connection between the pair is usually made with a spring-like molecule, such as a different protein (spectrin, actin)[148,149] or even DNA helices[150]. These components are then genetically encoded into cells and can yield a detectable FRET signal upon certain biological events that cause the two fluorophores to closely interact, depending on the design of the biosensor. A schematic of the various FRET biosensors strategies is shown in Figure 8.

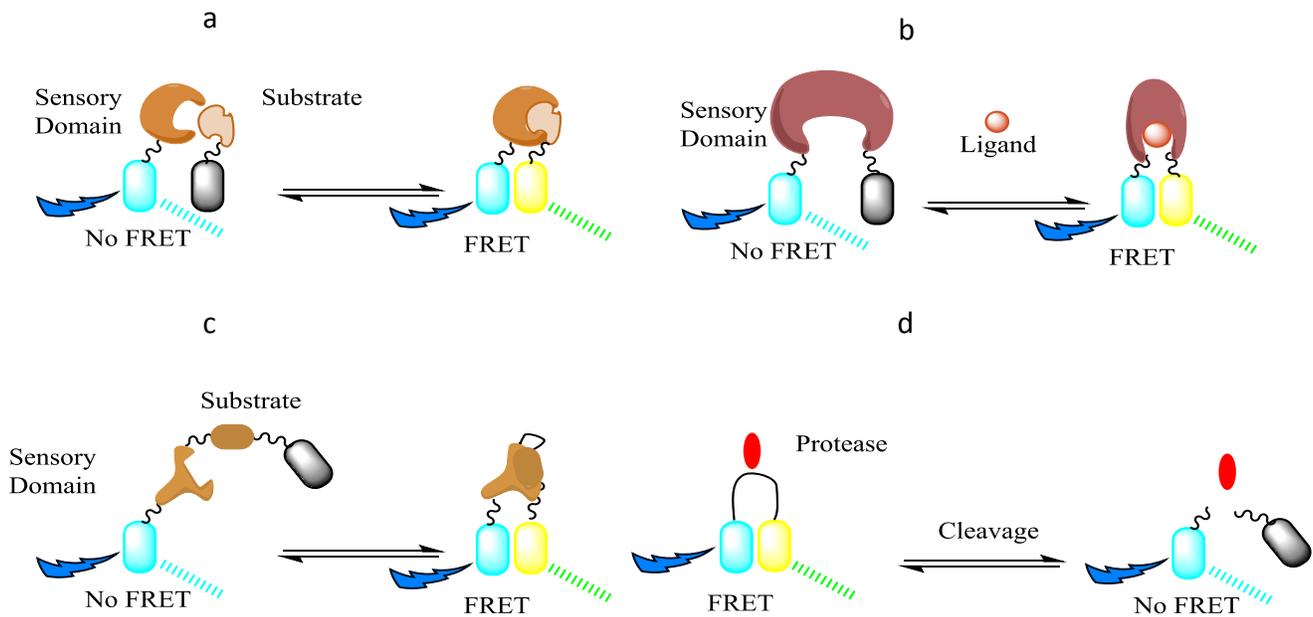

**Figure 8.** *FRET biosensors design strategies. The two fluorescent molecules are represented by the cyan and yellow cylinders, while the gray color indicated the lack of signal from the respective protein. (a) the donor molecule is attached to a sensory domain, while the acceptor is linked to the corresponding substrate, generating a FRET signal once they are binding; (b) both proteins are connected to a single sensory domain that changes its conformation upon binding to a particular ligand (c) both substrate and sensory domain are linked in a singular probe (d) an opposite strategy starts with the two proteins connected in close proximity to generate FRET signal, and detects the disappearance of it as the linker is cleaved/stretched in the biological media.*

FRET probes represent a tool that can provide molecular insight when combined with the types of assays described herein (scratch wound assays, migration assays, microfluidics-based methods etc.), enabling the understanding of underlying molecular mechanisms. For instance, Aoki et al. used a EKAREV-NLS expressing MDCK cells to study the effect of EKR waves, showing that the collective migration direction is facilitated by the propagation of EKR activation in 2D scratch-wound assays cultures[151]. Employing the same cell line, MDCK, but this time expressing a FRET vinculin TS tension biosensor, Abdellatef et al. studied the effect of the substrate on cell adhesion and its tension[152]. Moreover, Reffay et al. investigated the mechanisms of migration finger formations, showing a strong correlation between single-cell based RhoA activity gradients and local forces in the structure, by means of Rac1 and RhoA FRET probes[140]. Camona-Fontaine et al. [153] used FRET to probe for Rac1 activity in neural crest cells from zebrafish (in vivo) during embryogenesis, as they collectively migrate due to a proposed "coattraction" phenomenon. Using these probes in an in vivo assay provides an interesting possibility of observing factors that might be overlooked or oversimplified by in vitro model assays.

## IV. Computational Methods

### 1. Self-propelled particle (SPP) model

Self-propelled particle (SPP) was first introduced as a particular case of the "boids model" by Vicsek et al. in 1995 [154,155], where the particles align themselves with the average direction of neighboring particles. This model was introduced to investigate the transport and phase transition in non-equilibrium systems. Here, a particle within the system possesses a constant absolute velocity whose direction is determined as the average direction of motion of the neighboring particles within a searching radius *r* with some random perturbation added [154]. In particular, the velocity $\{v_i\}$ of the particles is identified at each time step, and the position of the *i*th particle is determined as:

$$x_i(t+1) = x_i(t) + v_i(t)\Delta t$$

Note that the velocity of a particle $v_i(t+1)$ is calculated to have an absolute value $v$ and a direction determined as the angle $\theta(t+1)$ as:

$$\theta(t+1) = \langle \theta(t) \rangle_r + \Delta\theta$$

where $\langle \theta(t) \rangle_r$ is the average direction of the velocities of neighboring particles within a circle of radius *r* surrounding the given particle (i.e. *i*th particle). Also, $\Delta\theta$ denotes a random number to represent the noise in the system. This model predicted that particles moved either in disordered or ordered motion depending on particle density (or cell packing fraction) and noise level.

Although this model can simulate collective cell migration, it has several disadvantages as the particles were simply modelled as points, and intercellular interaction was not considered. Researchers then extended and expanded this model to consider this interaction. Specifically, the intercellular force $F(r_i, r_j)$ between *i*th and *j*th cells is considered as piecewise linear force function that is a function of distance between two cells (Figure 9A) [59]. This force is repulsive if the distance is smaller than $R_{eq}$, whereas it is attractive when the distance is $R_{eq} \leq d_{ij} \leq R_0$, i.e.:

$$F(r_i, r_j) = e_{ij} \begin{cases} F_{rep} \dfrac{d_{ij} - R_{eq}}{R_{eq}}, & d_{ij} < R_{eq} \\ F_{adh} \dfrac{d_{ij} - R_{eq}}{R_0 - R_{eq}}, & R_{eq} \leq d_{ij} \leq R_0 \\ 0 & R_0 < d_{ij} \end{cases}$$

where $e_{ij} = \frac{r_i - r_j}{|r_i - r_j|}$, $d_{ij} = |r_i - r_j|$, $F_{rep}$ is the maximum repulsive force at $d_{ij} = 0$, and $F_{adh}$ is the maximum attractive force at $d_{ij} = R_0$. Moreover, skewing away from Vicsek's alignment to neighboring cells behavior, this model includes cells that align to the direction of the net force acting upon them.

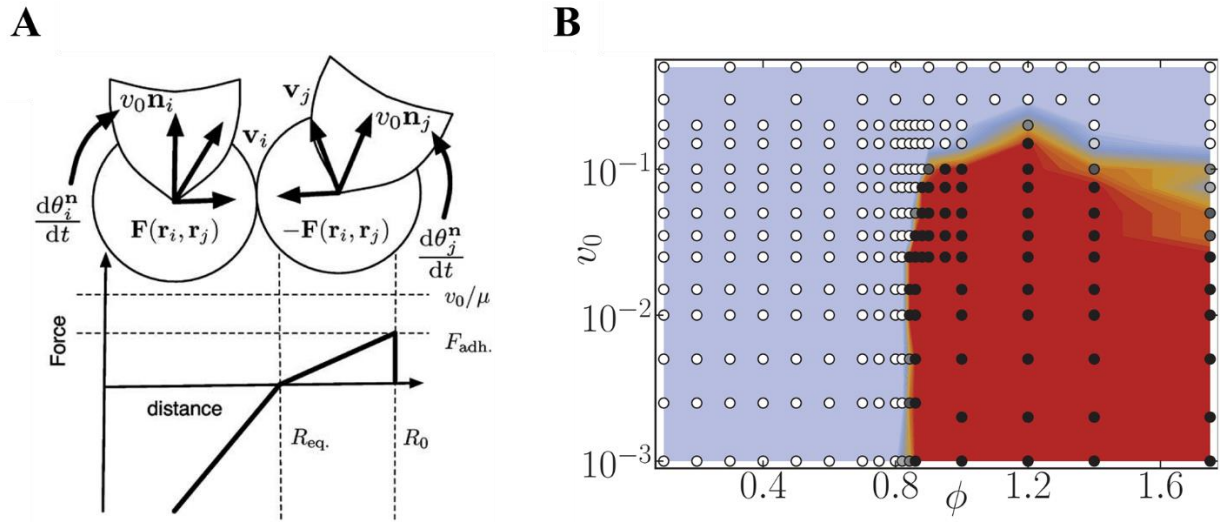

Figure 9. (A) Schematics showing intercellular forces as well as self-propelled velocity directions; Reprinted from [59]; (B) Phase diagram showing the transition from the liquid state (blue region) to the solid state (red region); Reproduced with permission from [156]. Copyright 2011 American Physical Society

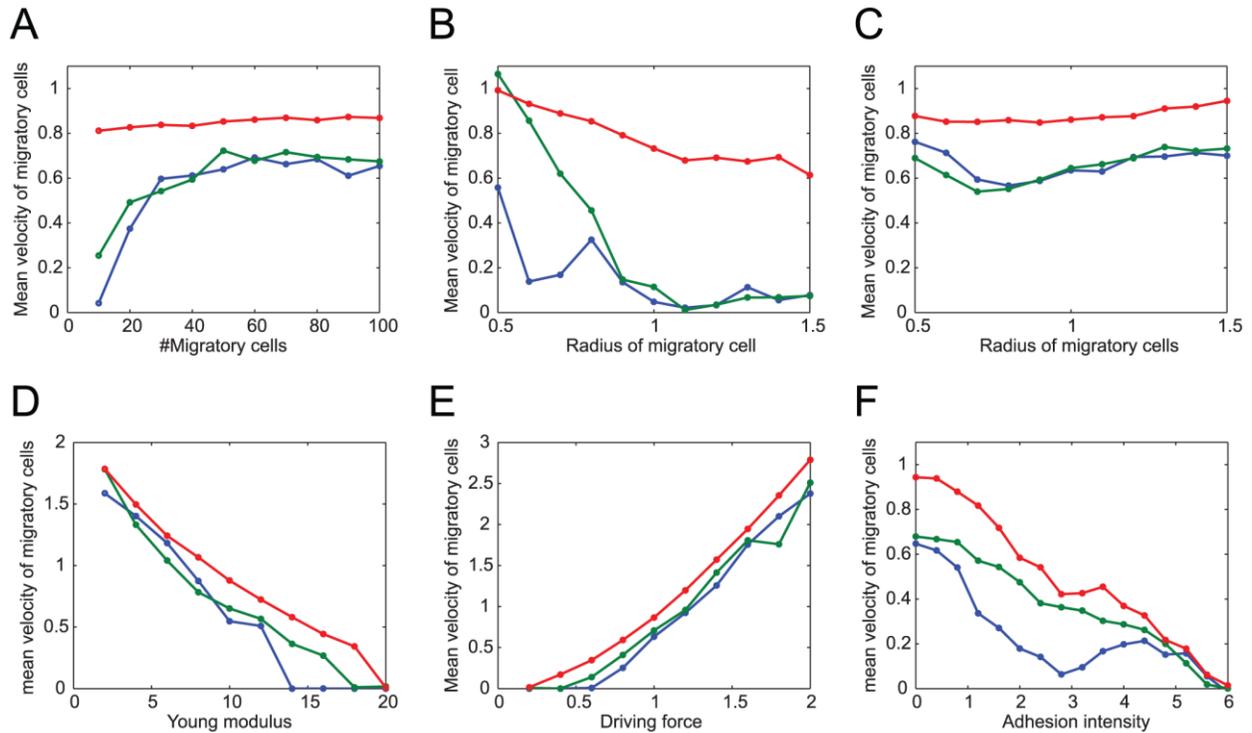

Figure 10. Plots of the collective (green), neutral (blue), and dispersive (red) migrations modes. The average migratory cell speeds are represented against (A) number of migratory cells, migratory cells' radius (B, C), migration driving force (D), and Young's modulus for all cells (E). In (F), an additional attractive force from cell adhesion is included. Reproduced under Creative Commons from[157]. Copyright 2011 Yamao et al.

This model was then further improved to investigate the migration behavior of cells [156–159]. The results revealed that cell packing fraction, moving velocity and noise level controlled whether the cells migrated as collective or dispersive behavior. The phase diagram was developed by Henkes et al., who reported that the cells were in liquid- or solid-like state depending on cell packing fraction and velocity (Figure 9B) [156]. Figure 9B clearly showed a transition from a liquid-like (or unjammed) state to a jammed phase at critical packing fraction of $\phi_c \approx 0.842$. Yamao et al. concluded that cells migrated as collective, neutral or dispersive behavior depending on the strength of noise from migratory cells vs. non-migratory ones. Their model used three main components in modelling cell behavior, namely repulsive forces between cells, the driving force of migratory cells together with the reactive forces of neighboring cells via adhesion, and the stochastic forces involved in a random walk. Their study includes the influence of such parameters as the number of migratory cells, their size, elasticity as well as the impact of the driving force and the adhesion to the environment (Figure 10) [157]. In 2014, Li and Sun utilized this model to study the coherent rotational motion of 2D confluent cell monolayer [160]. They stated that the rotational motion of the cell depends on the geometrical shapes and mechanical properties of the cells. They also concluded that mechanical coupling between cells is sufficient to explain this motion. This model has also been used to study the effects of interaction between stromal and cancerous cells, which is important in tumor growth and metastasis [161], on collective migration behavior[162]. The authors reported that stromal cell-cancerous cell interactions are sufficient to generate collective movement.

2. Vertex model

The SPP model above was able to simulate a jamming/un-jamming transition when cell density or packing fraction $\phi_c < 1$. However, the transition can also take place even in non-proliferating confluent biological tissues. In this case, the packing fraction $\phi$ is close to unity (i.e. there are no gaps between cells). To address this problem, vertex models [73,74,77,160,163–166], which have shown great potential and been extensively applied, were proposed. In the vertex model, each cell is represented by a polygon with several vertices. For a tissue containing *N* cells, the mechanical energy of the whole tissue is expressed as:

$$E = \sum_{i=1}^{N} E_i = \sum_{i=1}^{N} K_A (A_i - A_0)^2 + K_P (P_i - P_0)^2 \tag{4}$$

where $A_i$ and $P_i$ are the cross-sectional area and perimeter of the *i*th cell, respectively; $A_0$ and $P_0$ are the preferred cross-sectional area and perimeter, respectively; $K_A$ and $K_P$ are the area and perimeter moduli, respectively. The first term is an elastic term on cell area accounting for a previously apparently elastic behavior observed in cell layers[50,167,168]. The second term derives from two terms: the first one is quadratic in the cell perimeter to model the hypothesized active contractility of the actin-myosin subcellular cortex, and the second one is linear in the cell perimeter to simulate an effective line tension resulting from cell-cell adhesion and cortical tension. In addition, a dimensionless effect target shape index is defined as $p_0 = P_0/\sqrt{A_0}$, which the models have assumed to be constant across the whole tissue.

In this model, a critical value of $p_0 = p_0^* \approx 3.81$, at which the rigidity transition took place, was determined (Figure 11A) [73]. Below this critical value, the tissue behaved as an elastic solid-like material with finite shear modulus due to the cortical tension is superior to cell-cell adhesion. In this case, the cells had rounded shape and the energy barriers for cell rearrangement transitions are finite (Figure 11B). In contrast, above this value, the cells had elongated shape, cell-cell adhesion dominated, and cell rearrangements' energy barriers vanished (Figure 11C), resulting in zero rigidity and fluid-like behavior.

This model was then combined with the SPP model into a so-called self-propelled Voronoi (SPV) model [136,169]. This model identifies the cell using only the center $(r_i)$ of Voronoi cells in a Voronoi tessellation of space. For a tissue containing *N* cells, the total mechanical energy of the tissue capturing intercellular interactions is defined by equation (4). The effective mechanical force acting on *i*th cell is given as $\boldsymbol{F}_i = -\boldsymbol{\nabla}_i E$. In addition, cells can move due to self-propelled motility like that of SPP model. The cell center $\boldsymbol{r}_i$ is controlled by these forces using the overdamped equation of motion as follows:

$$\frac{d\boldsymbol{r}_i}{dt} = \mu \boldsymbol{F}_i + v_0 \hat{\boldsymbol{n}}_i$$

where $\mu$ is the motility; $v_0$ is the self-propulsion velocity; and $\hat{\boldsymbol{n}}_i$ is the polarity vector to $i$th cell.

In this model, Bi et al. used the effective self- diffusivity $D_{eff}$ as one of the criteria to identify glass transition where $D_{eff}$ was non-zero when the tissue behaved as fluid-like material. Here, $D_{eff}$ refers to the ratio between the total systems' self- diffusion coefficient and the self-diffusion coefficient of a single, isolated cell. The order parameter chosen this way is unitless, and can serve as an identifier of the liquid-like or solid-like state of the represented monolayer. Also, a unitless shape index of the tissue was defined as $q = \langle P/\sqrt{A}\rangle$, which was the averaged shape index of the cells. The authors claimed that in the model where cells were not motile ($v_0 = 0$), $q$ is constant ~3.81 when $p_0 < 3.81$, and $q$ was increasing linearly with $p_0$ when $p_0 > 3.81$. Interestingly, $q$ could also be used to determine the glass transition for all values of $v_0$. As shown in 12A, the cells were isotropic in the solid phase (blue squares) and anisotropic in the liquid phase (orange circles). In addition, the blue dash-line indicated the boundary defined $q = 3.81$. The authors developed a 3D phase diagram as shown in 12B to account for the effects of persistence time scale $1/D_r$ for the polarity vector $\hat{\boldsymbol{n}}$. The jamming transition in their system takes place when the tissue has low motility, cell have a low target shape index and low persistence time.

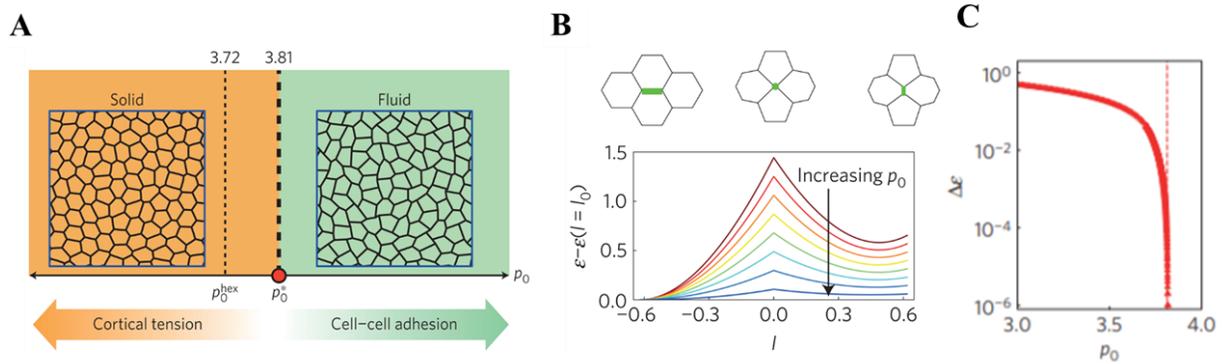

Figure 11. (A) Simple phase diagram showing the rigidity transition as a function of $p_0$. When $p_0 < p_0^* \approx 3.81$, the tissue was jammed (solid-like tissues), whereas the tissues was un-jammed (liquid-like tissue) when $p_0 > p_0^* \approx 3.81$; (B) Energy barriers for T1 transition as a function of edge length $l$ for different values of $p_0$, varying from 1.5 to 3.8 in equal increments; (C) the energy barrier height as a function of p₀, showing a sharp decrease with the approach of the critical value, p₀*=3.81, pointed by a vertical dashed line. Reproduced with permission from [73]. Copyright 2015 Springer Nature.

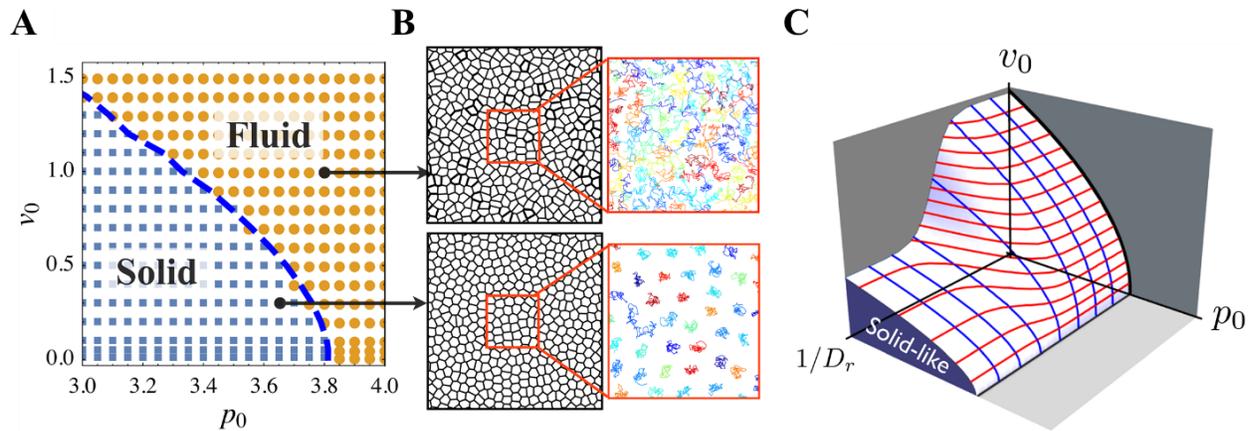

Figure 12. (A-B) Phase diagram of glass transition for confluent tissues as a function of cell motility $v_0$ and target shape index $p_0$ when $D_r = 1$; (C) 3D phase diagram of rigidity transition as a function of cell motility $v_0$, target shape index $p_0$ and persistence $1/D_r$. Reproduced with permission from [169]. Copyright 2016 Physical Review Society.

### 3. Cellular Potts model

A cellular Potts model is a stochastic model where each cell is represented as a subset of a lattice. This model has been used to study migration of cells through extracellular matrix [170] and through confluent layer of cells (Figure ) [171]. It can be observed that the migratory cells extended (yellow to red gradient) and squeezed between non-migratory cells by pushing them apart. In addition, Chiang et al. used this model to study the un-jamming transition in cell migration [172]. The dynamics of the cells calculated via Monte Carlo algorithm based on a Hamiltonian that included interfacial effects, approximate cell area conservation, and cell motility. Figure A and B show cell shape and trajectories for two different phases which were fluid-like (left panel in Figure A and B) and solid-like (right panel in Figure A and B) ones. It can be observed that cells had a more rounded shape in the solid-like phase where the cells were caged compared to an elongated shape in liquid-like phase where the cells were flowing. The authors also generated a phase diagram which is shown in Figure C demonstrating the transition from jammed solid to un-jammed phase depending on the interfacial energy $\alpha$ and strength of motility $P$. In this diagram, the blue circles and yellow square represented diffusive and sub-diffusive behavior of the cells. The red curve in this diagram indicated the critical shape index of the cells of $p_c = 4.9$. This value was slightly different with that identified in Vertex and SPV models (i.e. $p_c = 3.82$), which was probably due to discrepancies in cell morphologies in these models. These might arise from the infinite bending rigidity of the vertex's model edges, while CPM can resolve for more arbitrarily shaped cells.

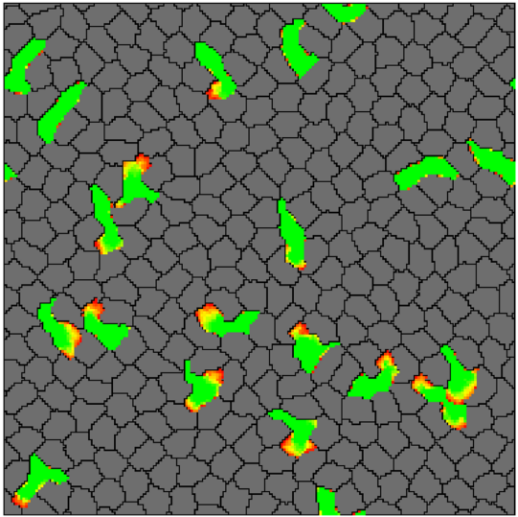

Figure 13. Simulation snapshot showing T cells (green) were squeezing between skin cells (gray) by pushing them apart. The yellow-red regions denote the protrusions extended by T cells. Reproduced with permission from [171]. Copyright 2015 PLoS.

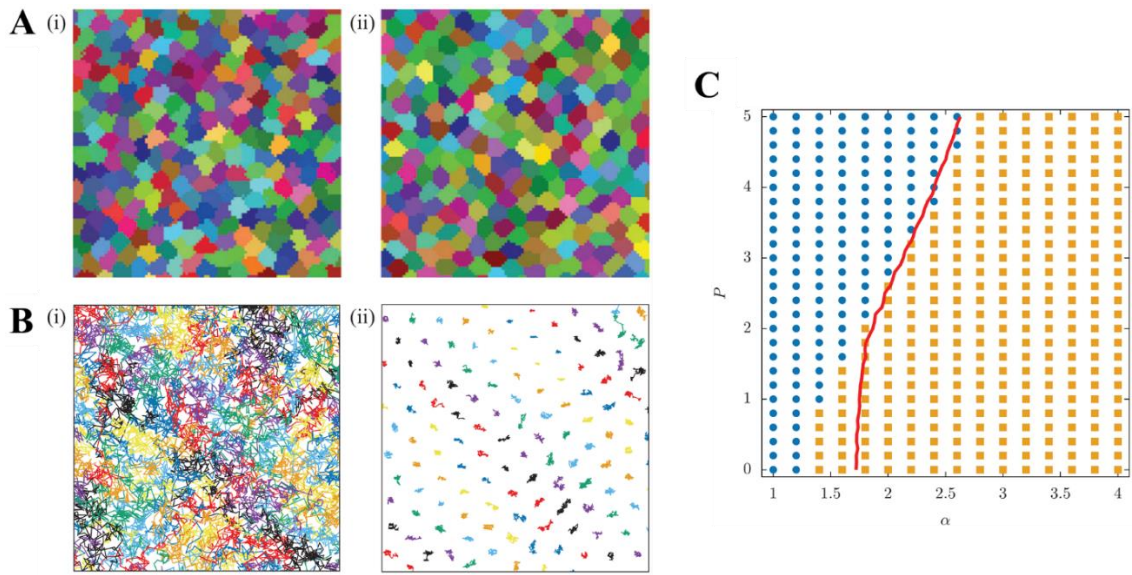

Figure 14. A snapshot of cell shapes (A) and trajectories of the cell centers for tissues in fluid (i) and solid (ii) state; (C) Phase diagram of rigidity transition as a function of the interfacial energy $\alpha$ and strength of motility $P$. The blue circles indicated diffusive behavior, whereas the orange squares indicated sub-diffusive behavior. The red line corresponds to the condition pairs under which the order parameter (or as previously described – shape index) was equal to its proposed threshold of 4.9. Reproduced with permission from [172]. Copyright 2016 EPL.

4. **Cellular Automaton (CA) model**

**CA model has been used extensively in theoretical biology. One of the first works applying this method in cancer cell invasion was introduced in 1985 by Duchting and Vogelsaenger** [173]**.**

This model was greatly improved since then to comprehensively investigate cancer cell invasion. For example, Jiao and Torquato developed a single-cell based CA model to study the growth dynamics and morphology of invasive solid tumors [174]. They first created non-overlapping spheres using a random sequential addition (RSA) process. The automaton cells were then created using Voronoi tessellation based on the centers of the spheres (Figure 15).

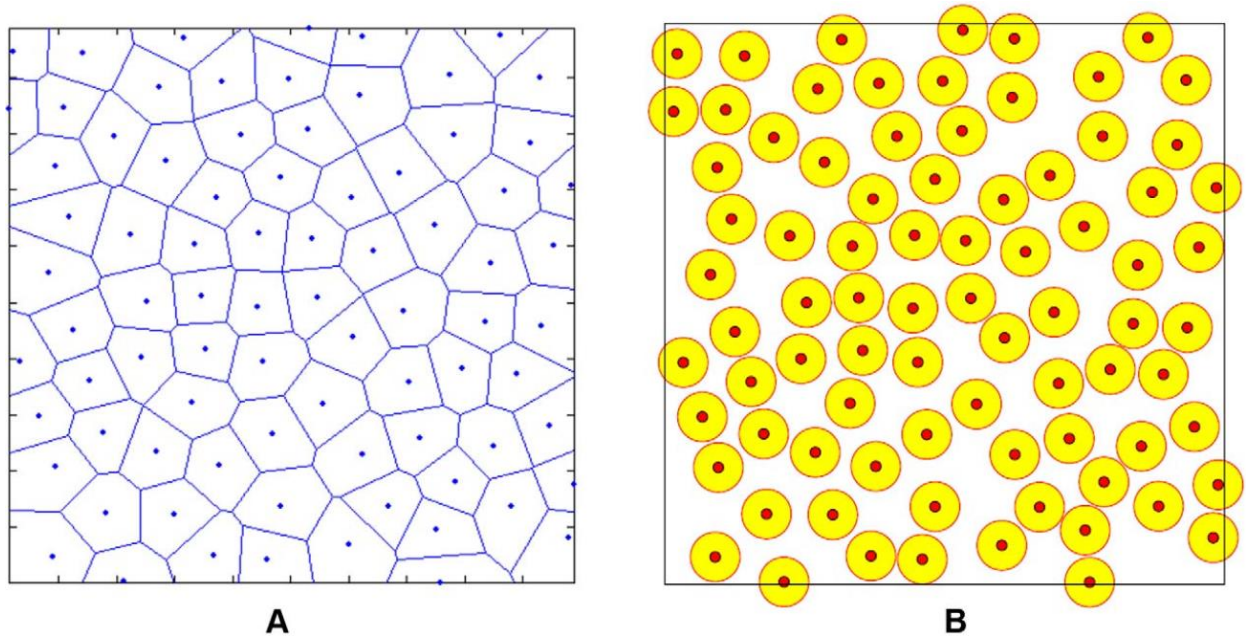

**Figure 15. Automaton cells were created using Voronoi tessellation (a) based on the centers of the spheres (b) which were created with RSA process.** Reproduced with permission from [174]. Copyright 2011 PLoS.

In this model, each automaton cell could be either ECM of a specify density, invasive, proliferative, quiescent, or necrotic cell. The microenvironment heterogeneity could easily be considered by varying the distributions of the ECM densities. An invasive cell would degrade the surrounding ECM and migrated from one automaton cell to another if the associated ECM in that cell has a "zero" density (i.e. fully degraded). The details rules of this CA model are illustrated in Figure 16. Using this model, the authors successfully captured the cancer cell invasion process by considering various tumor-host interactions (e.g., the mechanical interactions between tumor cells and microenvironment (ECM), degradation of ECM by the tumor cells and oxygen/nutrient gradient driven cell motions).

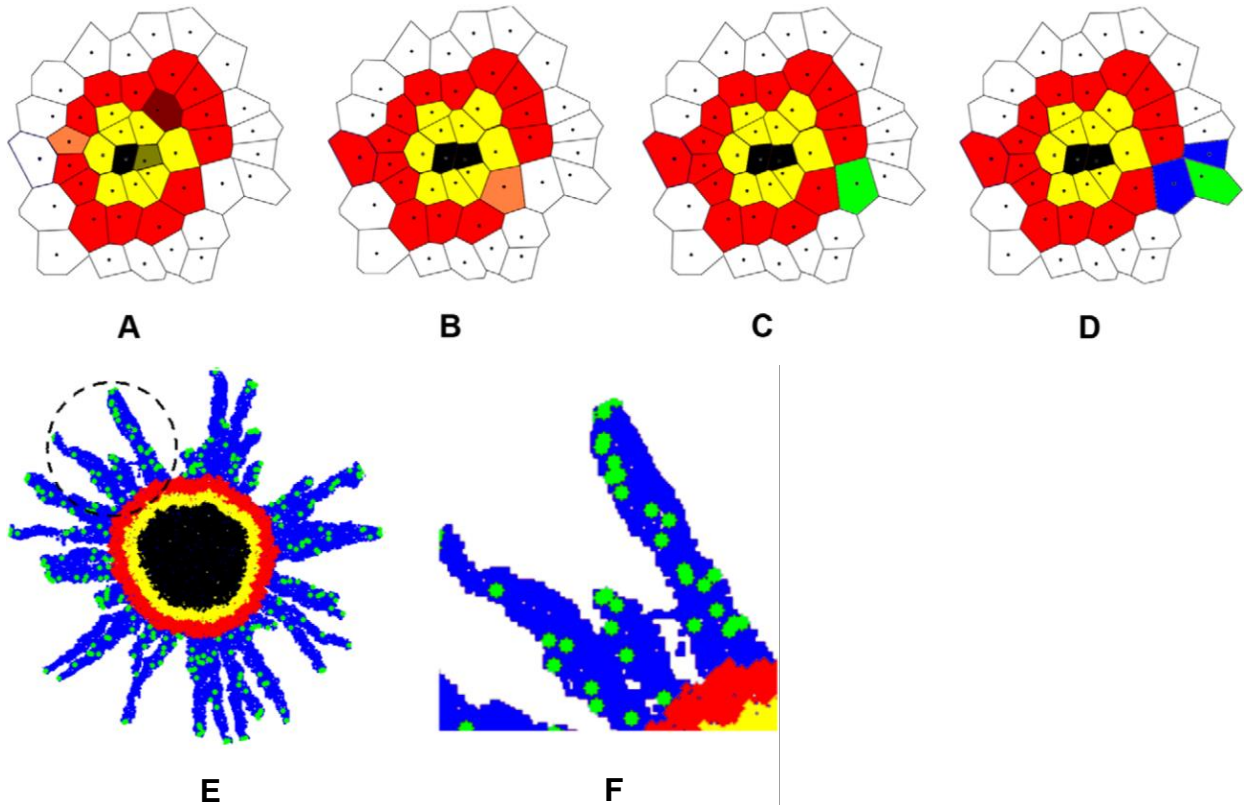

Figure 16. Black: necrotic cells; Yellow: quiescent cells; Red: proliferative cells; Green: invasive tumor cells; White: ECM; Blue: degraded ECM. (A) A proliferative cell (dark red) turns into a quiescent cell in panel (B) because it is too far away from the tumor edge. A quiescent cell (dark yellow) turns into a necrotic cell in panel (B) because it is too far away from the tumor edge. Another proliferative cell (light red) will create a daughter proliferative cell in panel (B). (B) A proliferative cell (light red) will create a mutant invasive daughter cell in panel (C). (D) The invasive cell degraded the surrounding ECM and moved to another automaton cell. (E) A snapshot of the CA model simulation showing cancer cell invasion process in a multicellular tumor spheroid at 24 hours after initialization. (F) A zoom-in view of the circled region in panel (E) [174].

This model was then combined with in vitro cell migration experiments to investigate the effects of ECM heterogeneity on cancer cell invasion [175]. The authors developed a heterogeneous ECM by creating a funnel-like Matrigel interface (Figure 17). The authors claimed that ECM heterogeneity is important in governing the collective cell invasive behaviors and therefore determining metastasis efficiency. Recently, the CA model was improved to investigate 3D tumor growth in heterogeneous environment under chemotherapy [176]. The authors concluded that constant dosing is more effective than periodic dosing in suppressing tumor growth. This 3D CA model could help researchers develop efficient tools for prognosis and to optimize cancer treatments.

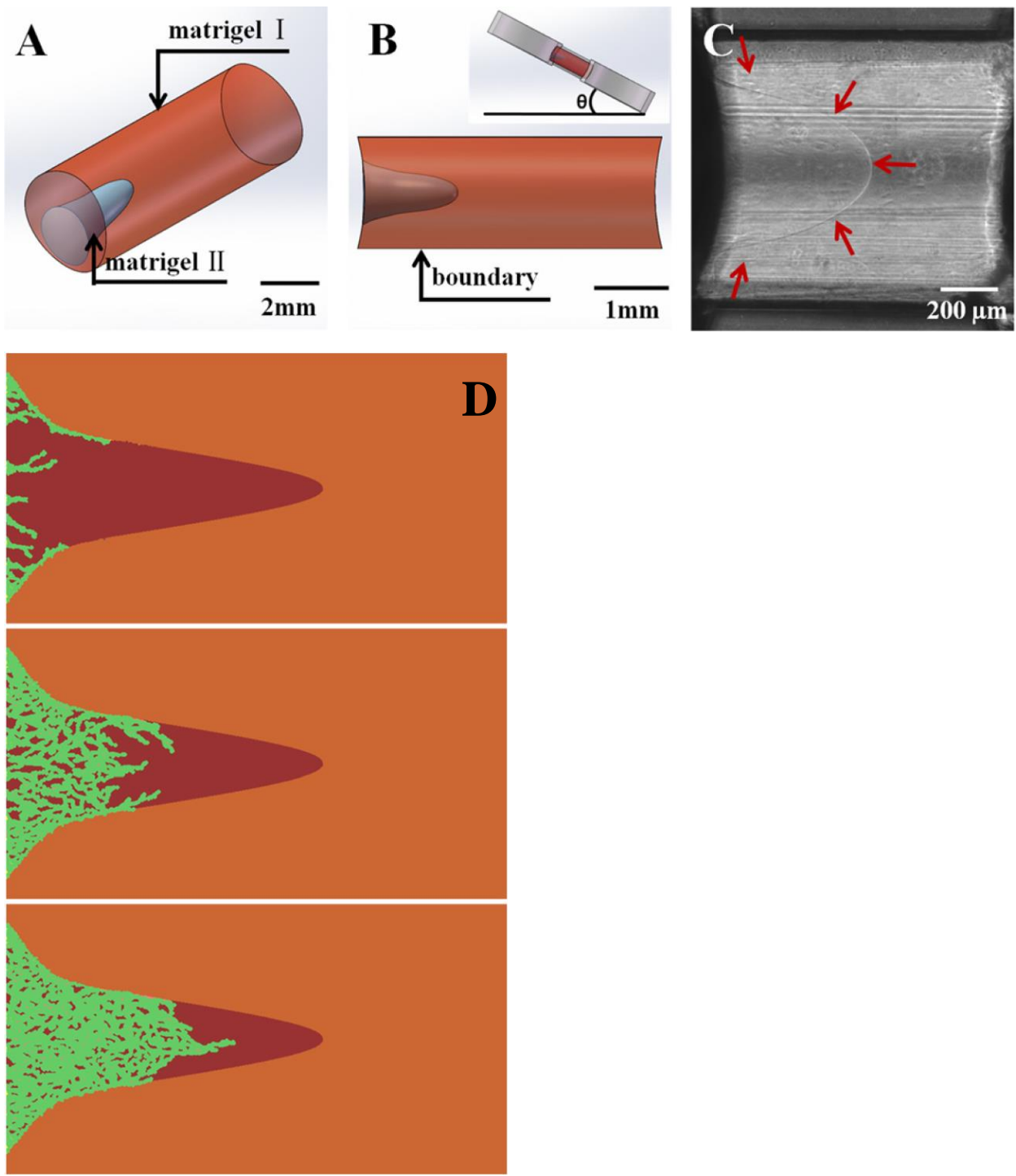

Figure 17. (A) and (B) Diagrams illustrate how funnel-like Matrigel interface is created with Matrigel I and II; (C) The gel interface is indicated by red arrows; (D) Snapshots of CA model simulation showing the collective migration pattern. Reproduced with permission from [175]. Copyright 2015 PLoS.

5. Finite Element (FE) model

The Finite Element (FE) model has been extensively used in cell and tissue mechanics studies [177–184]. Recently, this model was used to investigate cell-substrate adhesion as well as cell migration [185–188]. Wong and Tang developed a FEA model to investigate the effects of focal adhesion mechanical properties, substrate stiffness and intracellular stress on cell-matrix interaction during cell migration. The authors reported that cell-matrix traction had a biphasic relationship with respect to frictional coefficient, which could be used to identify focal adhesion properties as shown in Figure 18[185]. The simulation results revealed that both substrate stiffness and intracellular stress affected cell-matrix traction, but at different level. Particularly, traction increased in greater amounts when then intracellular stress was increased from 400 to 600 Pa compared to when substrate stiffness was increased from 0.5 to 100 kPa. Additionally, using a FEM-based mechano-chemical coupling model, Zhong et al. show that the competition of cell adhesion stability between the cell front and the cell rear drives individual cell migration[184]. While these studies focused mostly on single-cell migration dynamics, Zhao et al. proposed a so called dynamic cellular finite element model (DyCelFEM) to elucidate biochemical and mechanical cues in regulating cell migration and proliferation that include effects from neighbors in cell collectives (Figure 19) [186]. The authors concluded that improved directionality and persistence of cell migration was guided by biochemical cues, while mechanical cues played important roles in coordinating collective cell migration. It would be interesting to see whether the findings for single-cell migration determinants hold in the context of large-scale migration, when biochemical cues, geometrical factors and cell-environment interactions are treated in an integrated manner.

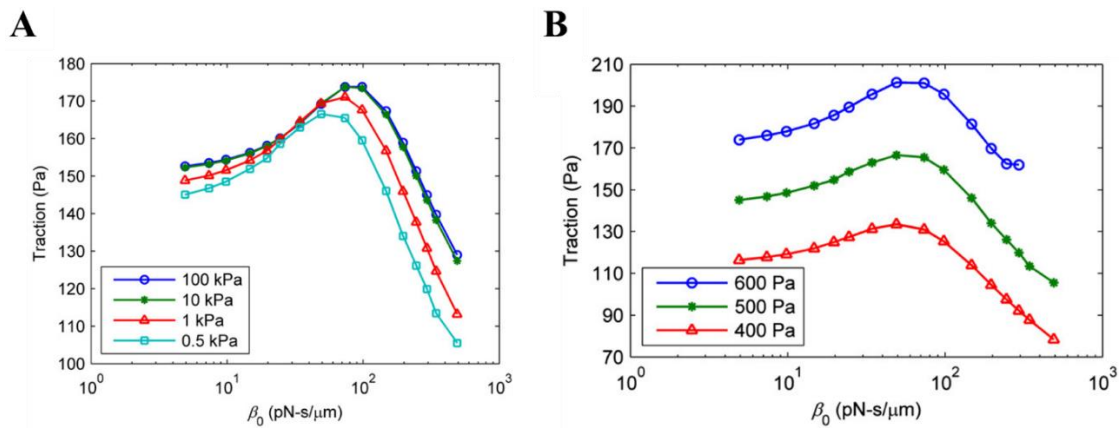

Figure 18. Maximum cell-substrate traction vs receptor-substrate friction coefficient plots for different substrate mechanical properties (A) and different maximum actomyosin stresses (B) Reproduced with permission from [185]. Copyright 2011 Elsevier Ltd.

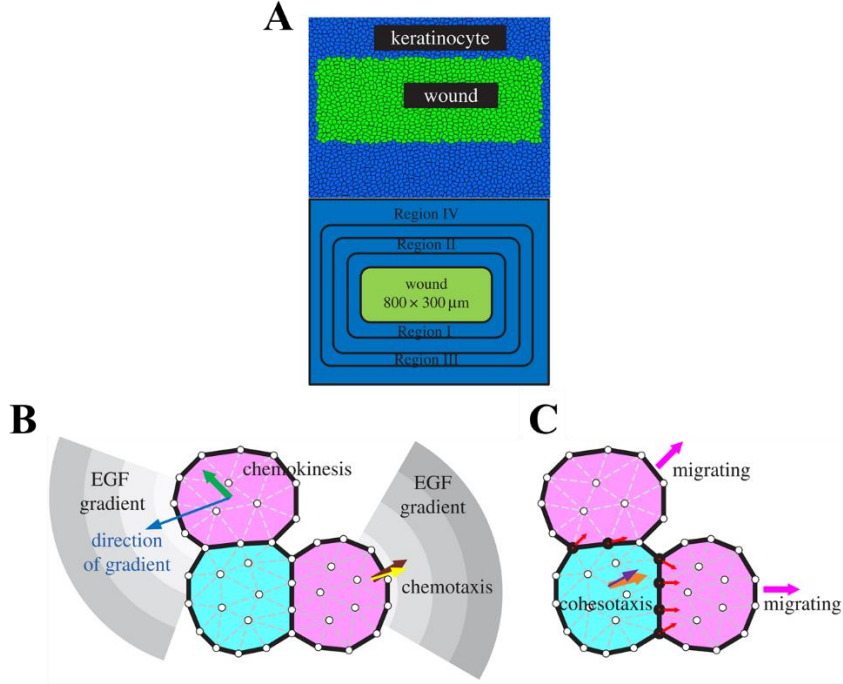

Figure 19. (A) Schematics of the skin wound tissue model; schematics showing how biochemical (B) and mechanical (C) cues regulate cell migration. Reproduced with permission from [186]. Copyright 2017 the Royal Society Publishing.

### 6. Particle-based model

Recently, a particle-based method (i.e. deformable particle model (DPM)) was developed to investigate the effect of cell packing density on jamming behavior of cell monolayers [189]. A cell is modeled as a polygon composed of a $N_V$ edges/particles. Adjacent particles are connected via linear springs with a spring constant of $k_l$ and equilibrium length of $l_0 = p_0/N_V$, where $p_0$ is the target perimeter of the cell/polygon. The total energy of DPM of a *N*-cell tissue is defined as:

$$U = \sum_{m=1}^{N} \sum_{i=1}^{N_V} \frac{k_l}{2}(l_{mi} - l_0)^2 + \sum_{m=1}^{N} \frac{k_a}{2}(a_m - a_0)^2 + U_{int}$$

where $l_{mi}$ is the length of *i*th edge of *m*th cell; $k_a$ is the area modulus which is corresponding to the cell's compressibility; $a_m$ and $a_0$ are the *m*th cell and preferred cross-sectional area. This is notably similar in form as the fundamental equation of the vertex model. In this equation, $U_{int}$ is the repulsive interactions between two adjacent cells given by:

$$U_{int} = \sum_{m=1}^{N} \sum_{n>m}^{N} \sum_{j=1}^{N_V} \sum_{k=1}^{N_V} \frac{k_r}{2}(\delta - |\boldsymbol{v}_{mj} - \boldsymbol{v}_{nk}|)^2 \times \Theta(\delta - |\boldsymbol{v}_{mj} - \boldsymbol{v}_{nk}|)$$

where $k_r$ is the strength of repulsive interaction; $\delta$ is the diameter of the disks centered at each of the vertices; $v_{mj}$ is the position vector of the jth particle in cell m; and $\Theta(.)$ is the Heaviside step function. The DPM showed that jamming took place at packing fraction of $\phi_{max} \approx 0.95 - 0.99$. Figure 20 shows the snapshots of cell shape at different packing fraction. The advantage of this model is that it can account for the cell membrane bending as well as its viscoelasticity, by integrating both an elastic and a viscous force between each two neighboring points on the membrane [190].

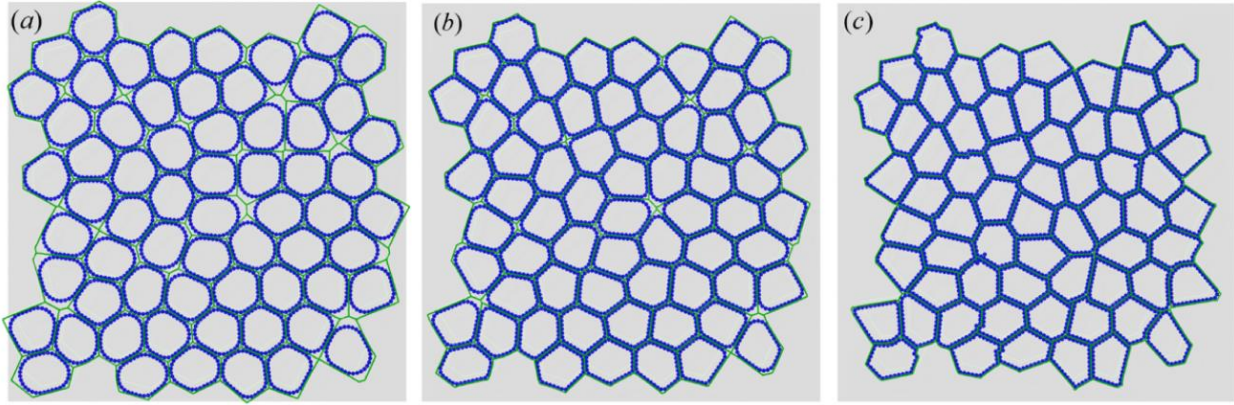

Figure 20. Snapshots of DPM results at different jammed packing fraction determined by a change in particle shape (described by the Asphericity factor, A), where (a) A=1.03, (b) A=1.08 and (c) A=1.16. Reproduced with permission from [189]. Copyright 2018 American Physical Society.

I.      Conclusions and Perspectives

Collective cell migration is a hallmark of events such as embryogenesis, wound healing or cancer tumor invasion[7]. Various studies at preclinical stages or using patient-derived samples have agreed on the fact that metastasis can be generated by clusters of cells rather than single cancer cells[191]. Moreover, the aggregation of tumor cells during blood circulation or at the distant organ site was shown to be highly inefficient[192], strongly supporting the hypothesis that clusters start as a collective cohort of cells from the primary tumor that migrate together to secondary sites and hence contributing significantly to the lethal nature of cancer.

As described in this review, numerous methods have been developed to study the biomechanical particularities and the implications on pathological progression of tumors. Experimental procedures, both in vitro and in vivo, alongside computational methods, have uncovered the puzzle pieces of a complex mechanism yet to be appropriately interconnected, which involves a variety of parameters such as cell-cell adhesions, cell-substrate interactions, microenvironment biomechanical behavior or cytoskeleton rearrangements that inform and regulate the collective cell migration behavior. However, there is still a way to go to unveil all the intricacies of such mechanism underlying collective cell migration. Experimentally, current limitations include the potential differences between the in vitro models (2 and 3 dimensional) and the in vivo realities, oftentimes difficult to address. Although experimental methods have been adapted for in vivo studies in the case of small organisms, observing collective cell migration triggering in mammals is still challenging[145,146]. On the other hand, computational models have not necessarily focused on collective migration in the case of cancer, even if basic principles and uncovered biomechanical factors still hold. Given the differences between cancer cells and their normal counterparts[28], as well as the particularities of the interactions within tumor stroma[162], some of the existing modelling strategies remain to be improved to account for such factors. Some advancements in the field have focused on the cell collective-ECM interactions during migration, showing that local stiffness anisotropy influences the migration direction[193] or that ECM fibers play an important role in long-distance mechano-signaling in both the single-cell[194] and multicellular cases[195,196]. These are promising strategies, especially in light of multiple experimental works showing a strong dependency of cell migration and ECM composition, stiffness or orientation.

From a clinical perspective, proposed methods of targeting collective cancer cell behavior include targeting leader cells[42,140] or reducing cell-cell communication[197]. However, even if technical hurdles of detection and genetic engineering could be overcome, previous results seem to shade the potential usability of such methods. For example, studies targeting cell-cell communication have shown that tumor aggressiveness increases while collective behavior decreases, indicating a tradeoff between collaboration and competitiveness[42,198]. Nonetheless, while the role of collective cell behavior is more evident in cancer metastasis, other features of tumors might be enhanced by it, including drug resistance and the support of stem cell niches[146]. Therefore, better understanding the mechanisms at play could allow for the right amount of tradeoffs as researchers develop new strategies to improve diagnosis or therapeutic efficacy.

**Acknowledgement**

Z. Chen acknowledges the support from the Branco Weiss - Society in Science Fellowship, administered by ETH Zürich. The research was in part supported by the National Cancer Institute of the National Institutes of Health under Award (No. U01CA202123). L. Liu acknowledges the support from the National Natural Science Foundation of China (Grant No. 11474345, No. 11674043).

**6,** 7 (2005).